\documentclass[twocolumn,
               aps,
               prd,   
               preprintnumbers,
               numbers,
               sort&compress,
               nofootinbib,
               showpacs,
               colorlinks,
               linkcolor=blue,   
               citecolor=blue]{revtex4-1}

\usepackage{graphicx,amsmath,amssymb,bm}
\usepackage{psfrag}
\usepackage{feynmp}
\usepackage{hyperref}
\usepackage{graphicx}
\usepackage{subcaption}
\usepackage{placeins}
\usepackage{xcolor} 
 \usepackage{lineno}

\def\<{\langle}
\def\>{\rangle}

\def\+{\dagger}

\def\U1A{U(1)$_{\rm A}$}

\def\ra{\rangle}
\def\la{\langle}

\newcommand{\be}{\begin{eqnarray}}
\newcommand{\ee}{\end{eqnarray}}
\newcommand{\beq}{\begin{equation}}
\newcommand{\eeq}{\end{equation}}
\newcommand{\exclude}[1]{}
\newcommand{\jcap}{JCAP}

\newcommand{\physrep}{Physics Reports}

\newcommand{\kmps}{\,\mathrm{km}\,\mathrm{s}^{-1}}

\begin{document}

\title{A {\bf Proposed} Network to Detect Axion Quark Nugget Dark Matter}
 
 \author{Xunyu Liang}
\email{xunyul@phas.ubc.ca}
\author{Egor Peshkov}
\email{e.peshkov@alumni.ubc.ca}
\author{Ludovic Van Waerbeke}
       \email{waerbeke@phas.ubc.ca}
 \author{Ariel Zhitnitsky}
\email{arz@phas.ubc.ca}
\affiliation{
 Department of Physics and Astronomy, University of British Columbia, Vancouver, Canada}
 
\label{firstpage}

\begin{abstract}
A network of synchronized detectors can increase the likelihood of discovering the QCD axion, within  the Axion Quark Nugget (AQN) dark matter model.  A similar network can also discriminate the X-rays emitted by the AQNs from the background signal. These networks can provide information on the directionality of the dark matter flux (if any), as well as its velocity distribution, and can therefore test the Standard Halo Model. We show that the optimal configuration to detect AQN-induced axions is a triangular network of stations 100 km apart. For X-rays, the optimal network is an array of tetrahedral units. 
   
\end{abstract}
\vspace{0.1in}


\maketitle

\section{ Introduction} 
 The Standard Halo Model (SHM) is locally characterized by an average dark matter (DM) density $\rho_{\rm DM}\approx 0.3~ {\rm GeV cm^{-3}}$ and a velocity distribution insensitive to  the environment, e.g. to the dynamics of the planets and the Sun (see \cite{Freese:1987wu}). Significant modifications of these local DM features would dramatically impact the entire field of direct DM searches. The SHM is consistent with the direct measurement of the average DM mass density from the latest GAIA data release, which finds $\rho_{\rm DM} = 0.6 \pm 0.4 ~{\rm GeV cm^{-3}}$ \citep{2019JCAP...04..026B}. However, this direct measurement of $\rho_{\rm DM}$ remains imprecise, even with the best kinematics measurement of the nearby stars available to date. Moreover, this measurement cannot rule out large variations of the local DM density at the scale of the Solar System \citep{2008arXiv0806.3767X,Lages_2013}. Furthermore, it has been suggested that the Sun and planets may play a key role in the local DM distribution \citep{Bertolucci:2017vgz,Zioutas:2020ndf, 2008JPhA...41G5201C,Nu_ez_Casti_eyra_2019}\footnote{In particular, \cite{Bertolucci:2017vgz} found strong correlations between the position of the planets and solar flare activities, and proposed that an ``invisible matter stream" is responsible for them. Furthermore, \cite{Zioutas:2020ndf}  claimed that stratospheric temperature anomalies cannot be explained by conventional atmospheric phenomena and are correlated with planetary positions.}. In addition, the Milky-Way merging history might locally alter the DM distribution, and even at the galactic scale, in nearby galaxies, some observed characteristics of DM are not satisfactorily explained by N-body simulations \citep{2020Univ....6..107D}.

In this work, we propose a framework that can
 test some  fundamental  assumptions of the SHM, such as the magnitude of the DM density and the DM velocity distribution in the local environment. These assumptions cannot be tested with a single  experiment, even a very large one. It requires a network of spatially distributed instruments. These detectors must be synchronized in order to study the directions of the DM particles hitting the Earth, and their velocity distribution.


 In a  recent proposal   \cite{Liang:2019lya,Budker:2019zka} it was suggested to use a broadband detection strategy  to search for axions within the context of the Axion Quark Nugget (AQN) dark matter   model. The main problem with  broadband  detection is the difficulty to distinguish between the AQN signal and large random  noise and spurious events, but  the dominant random background noise can be eliminated with the following observational strategy: 1- separate the DM signal from the larger noise by explicitly looking for annual and daily modulations,  which are specific to the former; 2- use synchronized detectors to discriminate between the DM signal and the larger noise, by looking at the time delays recorded by two or more nearby detectors, as these delays are unambiguously fixed  by the distances between the detectors. Developing such  a detection strategy is relevant to the search of any DM particles that leave a detectable trace, being neutrinos \cite{Zhitnitsky:2019tbh},  infrasonic/ seismic\footnote{Using a network of seismic detectors  around the globe, rather than a single instrument, allows to localize a seismic event with very high accuracy.} events \cite{Budker:2020mqk}, or X-rays. 
   
  We must emphasize that we will discuss correlated events that are synchronized but not coherent, for example distinct X-ray photons emitted in different directions at slightly different instants. Being emitted by the same AQN, they can be thought as clustering events which do not satisfy a conventional Poisson distribution. In contrast, noise spurious events are random and follow a conventional Poisson distribution, which allows to discriminate the AQN-induced signal from much larger noise events, provided one collects sufficient statistics of true signals. 
  
  Other networks searching for axions, such as The  Global Network of Optical Magnetometers for Exotic physics searches (GNOME)  \cite{2013arXiv1303.5524P,Afach:2018eze}  differ from our proposal. First, their search window is up to the kHz range, while the preferred value for the AQN axion mass corresponds to a frequency of  24 GHz. Second, they search for rare coherent single events, while we study correlated  clustering events on a much smaller scale, with a relatively high occurrence rate. 
   
The topic of the present study is two new elements, which have not been discussed previously in the literature:
   
   1. Three (or better four) neighbouring synchronized detectors allow to reconstruct the trajectory of an AQN, the source for the signals detected by all the stations.   This allows to study the DM flux distribution, including its velocity distribution and directionality.  
   
   2.  When an AQN propagates and annihilates in the Earth's atmosphere, it emits weakly coupled axions and neutrinos with very large mean free paths (as discussed in \cite{Budker:2019zka} and \cite{Zhitnitsky:2019tbh} respectively).  The AQN also emits X-rays, with a mean free path between 5 and 50 meters at sea level, depending on the energy, which interact strongly with the Earth environment. The study of cross-correlated X-ray signals can be more effective than current axion detectors \footnote{provided  the detectors we advocate  can be built and  be sensitive to relevant frequency bands. This is not a trivial assumption : the axion haloscopes known to be working, at present,  are based on cavity type detection, rather than on the broadband technology required by our study, see Section \ref{strategy} below for details.},    because  X-ray stations are easy to build and to operate, in comparison to more expensive and demanding axion detectors.  In order to discriminate AQN-induced X-rays  from background X-rays, one should arrange several synchronized detectors  separated by less than  50 meters, and study the correlated X-ray signals  emitted by  one and the same passing AQN. \footnote {A portion of the AQN-induced X-rays will be transferred  to sound waves, similar to conventional blasts resulting from bombs or large meteors, as recently argued in \cite{Budker:2020mqk} : the material vaporizes, rapidly expands, which eventually contributes to a small scale shock-wave (of order $10^{-3}{\rm cm}$). In our case, similar processes occur but at larger scales (of order 5 meters  or so \cite{Budker:2020mqk}). AQN-induced sound waves can propagate over  large distances (of order 100 kilometers \cite{Budker:2020mqk}), to contrast with X-rays with an attenuation length measured in meters. It means that one can study AQN-induced infrasounds and seismic correlated signals  using synchronized  stations. Such a network allows to study the DM distribution, its directionality and velocity distribution, but this is beyond the scope of this paper.}

In the next Section, we overview the basic ideas of the AQN framework, section \ref{strategy} is an overview of the known spectral features of the axion and X-ray emissions which play a key role in the present work. Our proposed network design is presented in Section \ref{network} and concluding remarks are discussed in Section \ref{conclusion}.

\section{The AQN framework}

First, we start with an overview of a novel mechanism for axion production, within the AQN framework,  as recently suggested in \cite{Fischer:2018niu,Liang:2018ecs,Lawson:2019cvy,Liang:2019lya}. This new   mechanism   always accompanies  the  two conventional  and well established axion  productions : by a misalignment mechanism ( when the cosmological field $\theta(t)$ oscillates and emits cold axions), or via the decay of topological objects\footnote{The AQN framework assumes   the conventional Peccei-Quinn resolution of the strong $\cal{CP}$  problem with the axion mass  assumed to be in the classical  window $m_a\in (10^{-6}-10^{-3})~ {\rm eV}$, see  original papers
\cite{1977PhRvD..16.1791P,1978PhRvL..40..223W,1978PhRvL..40..279W,KSVZ1,KSVZ2,DFSZ1,DFSZ2} and recent reviews \cite{vanBibber:2006rb, Asztalos:2006kz,Raffelt:2006cw,Sikivie:2009fv,Rosenberg:2015kxa,Marsh:2015xka,Graham:2015ouw,Battesti:2018bgc,Irastorza:2018dyq}. The AQN framework also assumes that inflation occurs during or after the PQ phase transition, such that one and the same misalignment angle $\theta_0$ occupies entire visible Universe.}.  
 
The AQN construction is in several aspects similar to the original quark nugget model suggested  by Witten \cite{Witten:1984rs} (see  \citep{Madsen:1998uh} for a review). It is a  type of ``cosmologically dark" DM, but not because nuggets are weakly interacting. AQN is a strongly-interacting DM candidate with a small cross-section-to-mass ratio, that reduces many observable consequences. 

There are two additional elements in the  AQN model compared to the old proposal \cite{Witten:1984rs,Madsen:1998uh}. First, there is an additional stabilization factor for the nuggets, provided by the axion domain walls (with the QCD substructure) which   are copiously produced  during the  QCD  transition. This extra pressure   alleviates a number of  problems with the original \cite{Witten:1984rs,Madsen:1998uh} nugget model.   Another feature of AQNs is that during the QCD transition, nuggets can be made of {\it matter} as well as {\it antimatter} . The direct consequence is that  the DM density, $\Omega_{\rm DM}$, and the baryonic matter density, $\Omega_{\rm visible}$, will automatically assume the  same order of magnitude (  $\Omega_{\rm DM}\sim \Omega_{\rm visible}$) , irrespective of the parameters of the model such as the axion mass $m_a$ or  misalignment angle $\theta_0$. 

We refer to the original papers   \cite{Liang:2016tqc,Ge:2017ttc,Ge:2017idw,Ge:2019voa} for questions  related to AQN formation, generation of the baryon asymmetry, and 
survival   pattern of AQNs in the hostile environment of the  early Universe. 
AQNs are characterized by a baryon charge  $B\in (10^{23}-10^{28})$ and a mass  $ M_N\simeq m_p|B|$. We emphasize that the AQN model is {\it  consistent with all presently available cosmological, astrophysical, satellite and ground-based constraints.  } This model offers a manifold of explanations for seemingly unrelated modern puzzles. It  alleviates tensions with observations, such as  the observed  excesses in galactic emission in various frequency bands. It may  resolve some longstanding cosmological puzzles, such as    the  ``Primordial Lithium Puzzle" \cite{Flambaum:2018ohm} and  the ``The Solar Corona Mystery"  \cite{Zhitnitsky:2017rop,Raza:2018gpb}.  It could explain ``impulsive radio events", recently observed in the quiet solar corona by the Murchison Widefield Array ( \cite{Ge:2020xvf}).  It may also explain  three unsolved puzzles: the annual modulation recorded by DAMA/LIBRA  (at a $9.5\sigma$ confidence level)  \cite{Zhitnitsky:2019tbh};   the X-ray seasonal variations recorded by the XMM-Newton observatory (at a $11 \sigma$ confidence level)    \cite{Ge:2020cho} ; the mysterious bursts observed by the Telescope Array \cite{Zhitnitsky:2020shd}. Finally, it could explain mysterious explosions, known as the "sky-quakes" \cite{Budker:2020mqk}, already mentioned above.

 We must emphasize that all the above studies  are based on  {\it  one and the same  } set of AQN parameters, even though they propose explanations for a wide variety of phenomena, in dramatically different environments where parameter differ by many orders of magnitude. In this work, we use again   {\it exactly the same set }  of parameters.

 For our estimates in the present work, we   need to  quote  the AQN hitting rate,  assuming a conventional dark matter density surrounding the Earth    $\rho_{\rm DM}\simeq 0.3\,{\rm  {GeV} {cm^{-3}}}$ . Assuming the SHM, one obtains  \cite{Lawson:2019cvy}:
 \be
\label{eq:D Nflux 3}
\frac{\langle\dot{N}\rangle}{4\pi R_\oplus^2}
&\simeq & \frac{4\cdot 10^{-2}}{\rm km^{2}~yr}
\left(\frac{\rho_{\rm DM}}{0.3{\rm \frac{GeV}{cm^3}}}\right)
 \left(\frac{\langle v_{\rm AQN} \rangle }{220~{\rm \frac{km}{s}}}\right) \left(\frac{10^{25}}{\langle B\rangle}\right),\nonumber \\
 \langle\dot{N}\rangle&\simeq & 0.67\,{\rm s}^{-1} \left(\frac{10^{25}}{\langle B\rangle}\right)\simeq  2.1\cdot 10^7 {\rm yr}^{-1} 
\left(\frac{10^{25}}{\langle B\rangle}\right)
\ee
where $\langle B\rangle \simeq 10^{25}$ is the average baryon charge for a given size distribution determined by the function 
$f(B)\propto B^{-\alpha}$, with $\alpha\approx (2-2.5)$ ( see    \cite{Zhitnitsky:2019tbh} for details and references). 
The averaging over all types of AQN-trajectories, with different masses $ M_N\simeq m_p|B|$,   different incident angles,   different initial velocities  and a variety of  models for size distribution,  does not modify this estimate noticeably.   The result (\ref{eq:D Nflux 3}) suggests that AQNs hit the Earth's surface with a frequency  of approximately  one a day   per   $(100 {\rm km)^2}$ area, which represent a typical ``small" event.     The hitting rate for large sized objects with $B\gg \langle B\rangle$  is suppressed by the distribution function $f(B)\propto B^{-\alpha}$ , while small sized objects with $B\lesssim \langle B\rangle$ represent more frequent but less intense events, according to the same distribution function $f(B)$. 
\exclude{It has been argued in   \cite{Budker:2020mqk} that  such a rare but very strong     event, with $B\simeq 10^{27}$, may  be responsible for  a sky-quake  which occurred on July 31st 2008  ( and was properly recorded by the dedicated Elginfield Infrasound Array (ELFO) near London, Ont., Canada). }

In  present work we study typical  events with $B  \simeq 10^{25}$. In the section \ref{network}, we will discuss how a network of synchronized detectors should be able to record the cross-correlations from such a single trans-passing nugget, and could allow to analyze the directionality and velocity distributions of the AQNs. But first, in the next section \ref{strategy}, we will overview the spectral features of the axion and related  X-ray emissions.

   \section{Spectral properties of the axion and X-ray emissions}\label{strategy}
 
In this section, we overview the spectral properties of the axion and X-ray emissions during  the passage of an antimatter AQN through the Earth, as these properties constrain the design of the instruments and the synchronized  network to be discussed in section \ref{network}.
 \subsection{Axion velocity distribution}\label{axion}
 We start our review  with AQN-induced axions. We represent the time-dependent axion flux 
  as follows \cite{Liang:2019lya}: 
\be
\label{flux}
\la E_a\ra \Phi^{\rm AQN}_a(t)\simeq 10^{14}A(t) \left[{\rm\frac{eV}{cm^2s}}\right], ~~~ \la E_a\ra\simeq 1.3 m_a, ~~
\ee
where the amplitude $A(t)$ is normalized to unity  when averaged over long period of time, i.e. $\la A (t)\ra =1$.
The background flux  in Eq. (\ref{flux})  is determined for axions emitted from deep Earth's underground using Monte Carlo simulations (as performed in \cite{Lawson:2019cvy,Liang:2019lya}) , for different size AQNs and different  trajectories. The atmospheric portion of the AQN's path was ignored in these simulations, because the density of the Earth's atmosphere is three orders of magnitude lower than the density of the planet itself. This should be contrasted  with the next subsection \ref{x-rays}
where the atmospheric portion of the path plays the dominant role. Here, the mean free path for X-rays in solids does not exceed 1 cm, while it is three orders of magnitude higher in the atmosphere.

There is  a number of time-dependent effects which influence $A(t)$. First, there is a well-established annual modulation  \cite{Freese:1987wu,Freese:2012xd} due to the changing position and velocity of the Earth relative to the Sun. It can expressed as :
 \begin{equation}
\label{eq:annual}
 A_{\rm (yr)}(t)= 1+\kappa_{\rm (yr)} \cos\left(\omega (t-t_0)\right)
\end{equation}
 where   $\omega=2\pi\,\rm yr^{-1}$ is the angular frequency of the annual modulation. The phase shift $\omega t_0$ corresponds to a maximum on July 1 and a minimum on December 1, assuming that the DM distribution is homogeneous and isotropic in the solar system environment.    
  Second, there is a daily modulation, unique to the AQN framework, that  can be described as \cite{Liang:2019lya} 
  \begin{equation}
\label{eq:daily}
 A_{\rm (d)}(t)=1+\kappa_{\rm (d)} \cos(\omega t-\phi_0)
\end{equation}
 where   $\omega=2\pi\,\rm day^{-1}$ is the angular frequency of the daily modulation. The phase shift  $\phi_0$ is similar to $\omega t_0$ in (\ref{eq:annual}). It  changes slowly with time due to the variations of the direction of  the DM wind  with respect  to the Earth's position and velocity. It can be assumed to be constant on the scale of days. In both cases, $A$ does not deviate from its average value by more than $(10-20)\%$. These modulations do not  constitute   the topic of the present study. 
  
 The main topic of our work corresponds to cases where $A(t)\gg 1$ for a short period of time $\tau$. These rare  bursts-like events are called ``local flashes'' in  Ref.\,\cite{Liang:2019lya}.  Typically, for $A\simeq 10^2$, $\tau \simeq$  1 second (see Table \ref{tab:local flashes}).  They result from AQNs hitting  the Earth and play a  key role in our study. Let's consider a network of detectors recording the cross-correlated signals emitted  by a trans-passing AQN.  Strong  signals  should  be observed by nearby detectors with a short time delays $\Delta t\sim d/v$ , where $d$ is the distance  between  detectors ( $d\sim 100$ km,  see \cite{Budker:2019zka}) and $v$ the velocity of the trans-passing AQN, $v\sim 10^{-3}c$, which must be distinguished from the velocity of the AQN-induced axion, $v_a$.  

 \begin{table} 
\captionsetup{justification=raggedright}
 	\caption{Estimations of local flashes for different $A$, as defined by (\ref{flux}).  The corresponding time duration $\tau$ and event rate  depend on $A$, which itself is determined by the shortest distance from the nugget's trajectory to the detector. The table is  adopted from \cite{Liang:2019lya}: } 
	\centering 
	\begin{tabular}{ccc}
		\hline \hline
		$A$ &  $\tau$ (time span) & event rate \\ 
		\hline 
		1 & 10 s & 0.3 $\rm min^{-1}$ \\ 
		$10$ & 3 s & 0.5 $\rm hr^{-1}$ \\ 
		$10^2$ & 1 s & 0.4 $\rm day^{-1}$ \\ 
		$10^3$ & 0.3 s & 5 $\rm yr^{-1}$ \\  
		$10^4$ & 0.1 s & 0.2 $\rm yr^{-1}$ \\  
		\hline 
	\end{tabular}
	\label{tab:local flashes}
\end{table}

 \begin{figure}
    \includegraphics[width=1\linewidth]{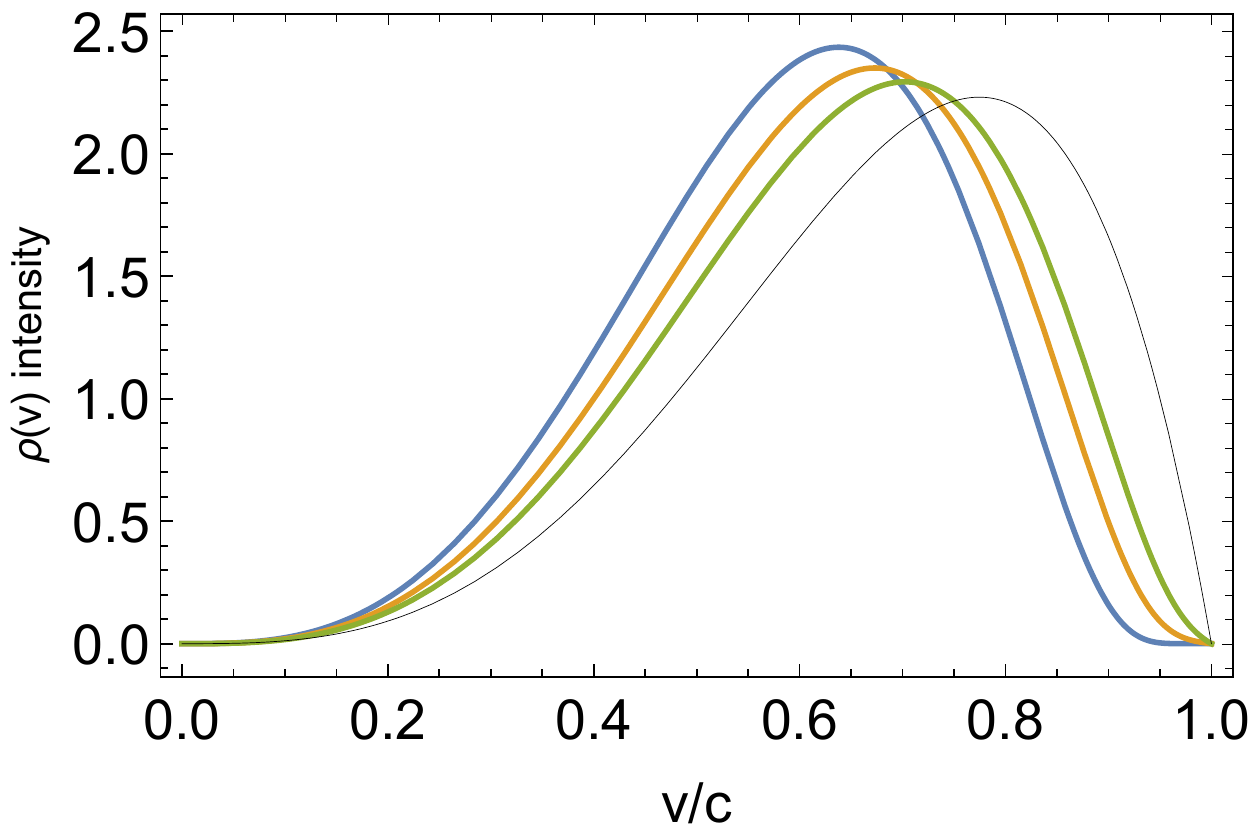}
    \caption{The normalized velocity distribution.  The plot is adopted from \cite{Liang:2018ecs}. The four different curves reflect the dependence on model parameters.}
    \label{spectrum}
\end{figure}
 
AQN-induced axions must be distinguished from  galactic axions, who have smaller velocities ( $\la v^{\rm gal}_a\ra \sim 10^{-3} c$)  : AQN-induced axions are relativistic ( $\la v_a\ra \sim 0.6 c$) while galactic axions are not ( see Fig. \ref{spectrum}) . The two types of axions have drastically different spectral features and these properties are important for reconstructing the trajectory of the trans-passing AQN. 

While the width of the spectrum shown on  Fig. \ref{spectrum}  does not influence the time delays $\Delta t$, this feature  drastically modifies   the energy of the photons induced  by the axion to photon conversion ($a\rightarrow \gamma$) in 
the   detectors.  Indeed, the axion's energy, for the dominant portion of the axions    in Fig. \ref{spectrum},  may vary from $E_a\simeq m_a$ to approximately $E_a\simeq 2m_a$\footnote{We use the natural units $\hbar=c=k_B=1$  throughout this paper.}. It implies that the  photons resulting from the $a\rightarrow \gamma$ conversion will be  distributed   with a bandwidth    $\Delta \nu \simeq  \nu $. This bandwidth should be contrasted with the narrow line (with  $\Delta \nu/\nu\sim 10^{-6}$ ) in the search for galactic axions,  in  cavity type detectors.  This is an essential difference between the two types of axions :
 { \it for AQN-induced axions,  the EM signal is very broad. } Therefore, one should use a broadband detection strategy  instead of the conventional cavity type detectors. 
 
   There is a number of proposals to design and build broadband   instruments :  QUAX \cite{Barbieri:2016vwg}, CASPEr \cite{JacksonKimball:2017elr}, ABRACADABRA \cite{Kahn:2016aff}, LC Circuit \cite{Sikivie:2013laa}, DM Radio \cite{Chaudhuri:2018rqn}, along with more recent   ideas    \cite{McAllister:2018ndu,Flower:2018qgb,Tobar:2020kmz,Tobar:2018arx}. All of these instruments 
   differ from cavity type instruments.  Some suggest to  use  different observables,  such as induced currents  in pickup loop (as advocated in   \cite{Kahn:2016aff,Sikivie:2013laa,Chaudhuri:2018rqn,McAllister:2018ndu,Flower:2018qgb,Tobar:2020kmz}).
While  there are presently no  broadband experiments operating in the window of interest ($ 10^{-6} {\rm eV} \lesssim m_a\lesssim 10^{-3} {\rm eV}$),  we do not foresee any fundamental  obstacles  to designing and building  such an instrument  in future.   In what follows, we assume that broadband axion detectors sensitive to a  $\Delta \nu \simeq  \nu $  distribution can be built and assembled in a network  (see section \ref{network}).

  \subsection{The X-ray emission and its spectral features}\label{x-rays}
  
  Here,  we overview the spectral characteristics of X-rays  emitted by the annihilation  of an AQN entering the Earth's atmosphere. In the previous section \ref{axion} , the dominant portion of the axions was produced deep underground,  while the atmospheric portion of the AQN's path was neglected. In the present subsection, on the contrary, the dominant portion of X-rays is emitted when the AQN propagates in the atmosphere, while the underground path can be ignored.

The spectrum of AQNs at low 
temperatures was analyzed in \cite{Forbes:2008uf} and was found to be
\be
  \label{eq:P}
  \frac{d{F}}{d{\omega}}(\omega) 
  &\simeq&
  \frac{1}{2}\int^{\infty}_{0}\!\!\!\!\! d{z}\;
  \frac{ d{Q}}{ d{\omega}}(\omega, z)\\
  &\simeq &
    \frac{4}{45}
  \frac{T^3\alpha^{5/2}}{\pi}\sqrt[4]{\frac{T}{m}}
  \left(1+\frac{\omega}{T}\right)e^{-\omega/T}h\left(\frac{\omega}{T}\right),\nonumber
\ee
where $Q(\omega, z)$ describes the emissivity per unit volume from the AQN's electrosphere. It is   characterized by a density  $n(z, T)$, where $z$ measures the distance from the quark core of the nugget. Function $h(x)$ in Eq.\,(\ref{eq:P}) is a slow varying  logarithmic   function  (explicitly computed in  \cite{Forbes:2008uf}). 
 
Integrating over $\omega$  gives  the total surface emissivity:
\begin{equation}
  \label{eq:P_t}
  F_{\text{tot}} = 
  \int^{\infty}_0\!\!\!\!\! d{\omega}\;
  \frac{d{F}}{d{\omega}}(\omega) 
  \simeq
  \frac{16}{3}
  \frac{T^4\alpha^{5/2}}{\pi}\sqrt[4]{\frac{T}{m}}\\
\end{equation}
  \begin{figure}
    \centering
    \includegraphics[width=1\linewidth]{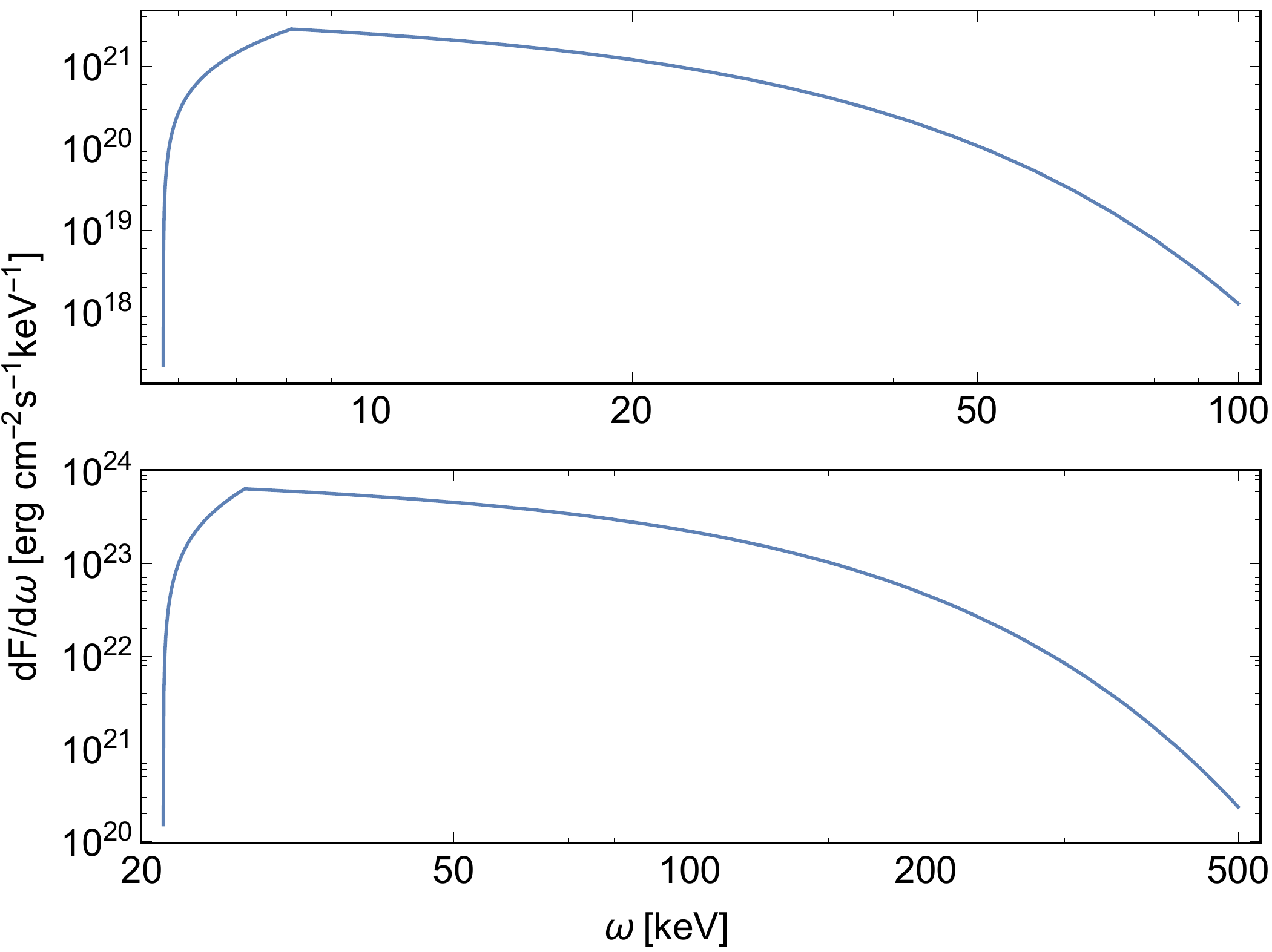}
    \caption{The spectral surface emissivity of an AQN with   the suppression effects at $\omega\ll T$.    The top plot  corresponds to  $T=10$ keV  while the bottom plot corresponds to $T=50 $ keV. Adopted from  \cite{Budker:2020mqk}.}
    \label{spectrum-1}
\end{figure}
where the temperature of the electrosphere\footnote{For an antimatter AQN, the electrosphere consists of positrons.}, for an AQN propagating in the atmosphere, is $(10-40)$ keV, depending on the altitude and  model-dependent parameters described in Appendix A (see \cite{Budker:2020mqk} for details). The corresponding spectrum is presented on Fig. \ref{spectrum-1}. 

The spectrum in Fig. \ref{spectrum-1} exhibits three important features.  First, it is almost flat in the region $\omega\lesssim T$, which is a direct manifestation of the soft Bremsstrahlung radiation (the emission of a photon with frequency $\omega$ is proportional to $d\omega/\omega$). Second, for large $\omega\gg T$,  the exponential suppression term, $\exp(-\omega/T)$, becomes the dominant component of the spectrum. Third, due to the function $h(x)$, the  emission is strongly suppressed at very small $\omega\simeq \omega_p\ll T$, where $\omega_p$ is the plasma frequency
   \begin{equation}
\label{eq:omega}
    \omega_p^2(z, T)=\frac{4\pi\alpha n(z, T)}{m}.
\end{equation}The drastic intensity drop  at small $\omega\ll T$ implies that visible light emitted by the AQN with $\omega\sim (1-10)$ eV  is  strongly suppressed in comparison to the X-ray emission. It implies that AQNs cannot be observed by optical monitoring. This low frequency suppression plays a key role in the   interpretation of sky-quakes. For example, \cite{Budker:2020mqk} could explain why the sky-quake event observed by ELFO  was not observed by all-sky cameras, and therefore was not a conventional meteoroid. 

Looking at sky-quakes, one might ask if  AQNs can be detected by seismic and infrasound instruments. Sky-quakes are identified with annihilations of AQNs  with a very large baryon charge (e.g. $B\simeq 10^{27}$ for the quake observed by ELFO). Such events only occur once every 10 years. According to (\ref{eq:D Nflux 3}), annihilations of AQNs with $B\simeq 10^{25}$ occur approximately once a day in area $\sim (100~ {\rm km})^2 $.  Conventional seismic  or infrasound instruments are not sufficiently sensitive to record such numerous but weak events. For those, we propose to use a network of  X-ray detectors, described in the next section.

\section{Trajectory and velocity reconstruction.}
\label{network}  

\begin{figure}[h]
	\centering
	\includegraphics[width=\linewidth]{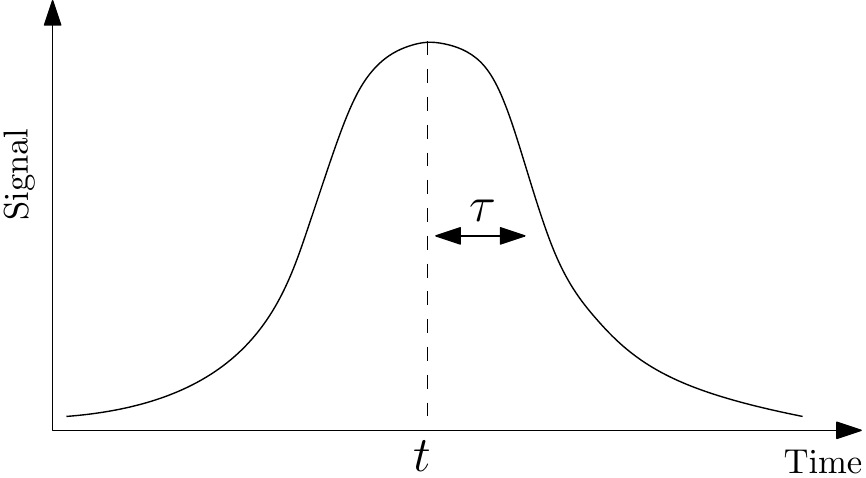}
	\caption{Signal (axion or X-ray) of a local flash observed by a detector. The burst-like signal peaks at time $t$, with a bandwidth $\tau$.}
	\label{fig:signals}
\end{figure}

When an event is detected at time $t$, the instrument records a burst-like signal with bandwidth $\tau$, as shown in Fig. \ref{fig:signals}. In a synchronized network of detectors, the time delay $(t_i-t_j)$ between two nearby stations located at positions $\mathbf{R}_i$ and $\mathbf{R}_j$ is
\begin{equation}
\label{eq:t_i-t_j}
t_i-t_j
=\frac{(\mathbf{R}_i-\mathbf{R}_j)
	\cdot\mathbf{v}}{v^2}\,.
\end{equation}
Here, the trajectory of an AQN is assumed to be linear \cite{Lawson:2019cvy} with a constant velocity $\mathbf{v}$ over the short duration
\begin{equation}
\label{eq:tau}
\tau
\equiv\frac{d}{v}
=0.3{\rm\,s}
\left(\frac{d}{100{\rm\,km}}\right)
\left(\frac{300\rm\kmps}{v}\right)\,
\end{equation}
to be compared to the typical time ($\sim10$ s) required by an AQN to traverse the Earth. Here, $d$ is the closest distance from a detector to the detected AQN. When detecting axions, the optimal distance is $d\lesssim100$ km, otherwise the benefit of local flashes diminishes \cite{Budker:2019zka}. When detecting X-rays, $d\lesssim$ 50 m is no more than its attenuation length at the sea level.

Depending on the radiation source, instruments may record more features than just $t_i$ and $\tau_i$, e.g. the incident direction of the flux. This is beneficial but not essential, as we will see $t_i$ and $\tau_i$ are sufficient to determine the trajectory of an AQN. In this work, we assume that $t_i$ and $\tau_i$ are the only data accessible, while additional detected information are devoted to improving the precision in $t_i$ and $\tau_i$.

\subsection{Reconstruction from a synchronized network}
\label{subsec:Reconstruction from a synchronized network}
In a synchronized network, it is convenient to choose one station (labeled as ``station 1'') as the reference location and as the starting time:
\begin{equation}
\label{eq:bold R_1}
\mathbf{R}_1=\mathbf{0}\,,\quad
t_1 = 0\,.
\end{equation}
Hence, the trajectory of an AQN is  
\begin{equation}
\label{eq:bold r(t)}
\mathbf{r}(t)=\mathbf{v}t+\mathbf{d}_1\,,
\end{equation}
where $\mathbf{d}_1$ is the closest vector distance from the $\mathbf{R}_1$ to $\mathbf{r}(t)$.

\begin{figure}[h]
	\centering
	\includegraphics[width=0.9\linewidth]{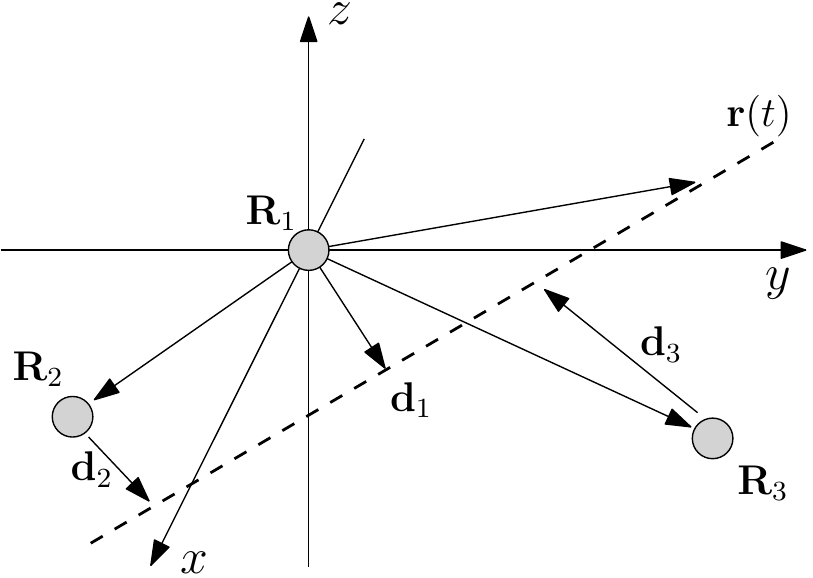}
	\caption{Coordinate system used in Eqs. \eqref{eqs:bold d_1 cdot bold v etc}. An AQN passes nearby three stations (grey circles) located at $\mathbf{R}_i$ ($i=1,2,3$), along a linear trajectory $\mathbf{r}(t)$ (dashed). The distance from each station to $\mathbf{r}(t)$ is denoted by $\mathbf{d}_i$ ($i=1,2,3$) respectively.}
	\label{fig:coord}
\end{figure}

In the trajectory \eqref{eq:bold r(t)}, there are 6 scalar unknowns: the components of $\mathbf{v}$ and $\mathbf{d}_1$. At least six independent scalar equations are therefore needed to reconstruct the trajectory \eqref{eq:bold r(t)}. For a network of three synchronized stations (see Fig. \ref{fig:coord}), one can show that:
\begin{subequations}
	\label{eqs:bold d_1 cdot bold v etc}
	\begin{equation}
	\label{eq:bold d_1 cdot bold v}
	\mathbf{d}_1\cdot\mathbf{v}=0\,,
	\end{equation}
	\begin{equation}
	\label{eq:d_1}
	d_1=\tau_1 v\,,
	\end{equation}
	\begin{equation}
	\label{eq:bold R_2 cdot bold v}
	\mathbf{R}_2\cdot\mathbf{v}=v^2t_2\,,
	\end{equation}
	\begin{equation}
	\label{eq:bold R_3 cdot bold v}
	\mathbf{R}_3\cdot\mathbf{v}=v^2t_3\,,
	\end{equation}
	\begin{equation}
	\label{eq:abs bold r(t_2)}
	\left(\frac{\tau_2^2+t_2^2}{\tau_1^2}-1\right)d_1^2
	+2\mathbf{R}_2\cdot\mathbf{d}_1
	=R_2^2\,,
	\end{equation}
	\begin{equation}
	\label{eq:abs bold r(t_3)}
	\left(\frac{\tau_3^2+t_3^2}{\tau_1^2}-1\right)d_1^2
	+2\mathbf{R}_3\cdot\mathbf{d}_1
	=R_3^2\,.
	\end{equation}
\end{subequations}
Eq. \eqref{eq:bold d_1 cdot bold v} is the orthogonality condition by definition of $\mathbf{d}_1$. Eq. \eqref{eq:d_1} is the definition \eqref{eq:tau} of bandwidth. Eqs. \eqref{eq:bold R_2 cdot bold v} and \eqref{eq:bold R_3 cdot bold v} are identical to Eq. \eqref{eq:t_i-t_j} in the reference frame  \eqref{eq:bold R_1}. Eqs. \eqref{eq:abs bold r(t_2)} and \eqref{eq:abs bold r(t_3)} are obtained from the definition of distances
\begin{equation}
	\label{eq:abs bold r(t_i)}
	|\mathbf{r}(t_i)-\mathbf{R}_i|
	=d_i
	=\frac{\tau_i}{\tau_1}d_1\,,\quad (i=2,3)\,,
	\end{equation}
and substituting into Eqs. \eqref{eq:bold R_2 cdot bold v} and \eqref{eq:bold R_3 cdot bold v} respectively.

These six equations Eqs. \eqref{eqs:bold d_1 cdot bold v etc} have no unique solution because they consist in an implicit  $\mathbb{Z}_2\times\mathbb{Z}_2$ symmetry, related to reflection and (modified) time reversal (see Appendix \ref{app:Degenerate solutions in Eqs.} for details). Consequently, for a given observation Eqs. \eqref{eqs:bold d_1 cdot bold v etc} yields four degenerate solutions, one of which being the true trajectory \eqref{eq:bold r(t)}. To eliminate this degeneracy, one can either introduce additional physical constraints or add more detectors, as discussed in subsection \ref{subsec:Optimal configurations of a synchronized network}.

\subsection{Optimal separation distance between stations}
\label{subsec:subsec:Sensitivity constraint of time measurement}
In a network of synchronized detectors, the separation distance $\Delta R$ between two neighboring stations should not be greater than an effective distance $d$:
\begin{equation}
\label{eq:Delta R lesseq}
\Delta R\lesssim d\,,
\end{equation}
where $d\simeq100{\rm\,km}\,$for axions and $d\simeq50{\rm\,m}\,$ for X-rays, otherwise the event rate of correlated signals diminishes due to the sharp reduction in their shared cross section \cite{Budker:2019zka}. On the other hand, $\Delta R$ cannot be too small because it is inversely proportional to the relative precision in the measurement of the time delay, $t_i$. Hence, optimal performance is achieved for a narrow range of $\Delta R$, which we will discuss below.

To estimate how the trajectory \eqref{eq:bold r(t)} is affected by the input data $t_i$ and $\tau_i$, we perturb Eqs. \eqref{eqs:bold d_1 cdot bold v etc}. We find a qualitative relation
\begin{equation}
\label{eq:delta v v}
\frac{|\delta \mathbf{v}|}{v}
\sim\frac{|\delta \mathbf{d}_1|}{d_1}
\sim\frac{\delta t_i}{t_i}
\sim\frac{\delta\tau_i}{\tau_i}\,,\quad
(i=1,2,3)\,,
\end{equation}
where, according to \eqref{eq:bold R_1},  $\delta t_1=0$. Only the uncertainty in $\mathbf{v}$ matters, because $d_1$ [of order 100 km (resp. 50 m) for axion (resp. X-rays)] is much smaller than the size of the Earth. A small perturbation $\delta d_1$  corresponds to a negligible shift of the trajectories at the Earth surface. Presumably, $|\delta\mathbf{v}|$ can be expressed in the empirical form of independent noise:
\begin{equation}
\label{eq:abs delta bold v v}
\frac{|\delta\mathbf{v}|}{v}
\leq C \sum_{i=1}^{3}
\sqrt{\left(\frac{\delta t_i}{t_i}\right)^2
	+\left(\frac{\delta\tau_i}{\tau_i}\right)^2},
\end{equation}
where $C$ is a dimensionless factor of order one. Monte Carlo simulation suggests $C\simeq 1.2$ for 95\% of samples (see Appendix \ref{subapp:Sensitivity to time measurement}). From Fig. \ref{fig:signals}, we expect $\delta t$ to be proportional to the bandwidth $\tau_i$:
\begin{equation}
\label{eq:delta t}
\delta t_i=\varepsilon\,\tau_i\,,
\end{equation}
where $\varepsilon\in(0,1)$ relates to the efficiency of measurement. For a moderately large $\varepsilon\gtrsim0.1$, it corresponds to a low signal-to-noise ratio. This is particularly true for axion detection. In this case, the $\delta t_i/t_i$ terms dominate inequality \eqref{eq:abs delta bold v v}:
\begin{equation}
\label{eq:langle abs delta v v rangle}
\begin{aligned}
\langle\frac{|\delta\mathbf{v}|}{v}\rangle
&\lesssim \sqrt{2}C\varepsilon\frac{d}{\Delta R}
\langle(\mathbf{\hat{R}}\cdot\mathbf{\hat{v}})^{-1}\rangle\,,\\
\langle(\mathbf{\hat{R}}\cdot\mathbf{\hat{v}})^{-1}\rangle
&\equiv\left<\sqrt{
	(\mathbf{\hat{R}}_2\cdot\mathbf{\hat{v}})^{-2}
	+(\mathbf{\hat{R}}_3\cdot\mathbf{\hat{v}})^{-2}
}\right>\,,
\end{aligned}
\end{equation}
where definitions \eqref{eq:t_i-t_j} and \eqref{eq:tau} were substituted. $\Delta R$ is defined as the average separation distance between two nearest stations, and the root mean square $\langle(\mathbf{\hat{R}}\cdot\mathbf{\hat{v}})^{-1}\rangle\simeq12$ from Monte Carlo simulation for 95\% of samples (see Appendix \ref{subapp:Sensitivity to time measurement}). Precision of measurement can be enhanced by installing more than one detectors in a station. Assuming there is a total of $N$ detectors in the synchronized network, to ensure that $|\delta\mathbf{v}|/v$ is sufficiently small, the constraint on nearest separation distance is
\begin{equation}
\label{eq:Delta R}
\begin{aligned}
\Delta R
&\gtrsim\sqrt{\frac{2}{N}}C\varepsilon
d\langle(\mathbf{\hat{R}}\cdot\mathbf{\hat{v}})^{-1}\rangle  \\
&=1.0\,d
\left(\frac{C}{1.2}\right)
\left(\frac{\varepsilon}{0.1}\right)
\left(\frac{
	\langle(\mathbf{\hat{R}}\cdot\mathbf{\hat{v}})^{-1}\rangle
}{
	12}\right)
\left(\frac{4}{N}\right)^{1/2}\,.
\end{aligned}
\end{equation}
$N$ must be {\bf at least} 4 (or 3 with additional physical constraints) to obtain a unique solution for Eqs. \eqref{eqs:bold d_1 cdot bold v etc}, as discussed in the previous subsection. Hence, even for a moderately low signal-to-noise ratio ($\varepsilon\gtrsim0.1$), three or four stations are sufficient to measure trajectories, provided their closest separation distance is of order $d$. 

To summarize, constraints \eqref{eq:Delta R lesseq} and \eqref{eq:Delta R} imply that the optimal value for $\Delta R$ is of order $d$, such that both the event rate and the precision of measurement are maximized.

\subsection{Optimal configurations of a synchronized network}
\label{subsec:Optimal configurations of a synchronized network}
The optimal configuration of a synchronized network depends on the type of radiation. For axions, a triangular network is more cost effectiveness because it has the highest ratio event rate per station, especially when operating on an extensible global network. If uniqueness of solution is a priority, a square network is better for axion detection. Another option is to construct a dense network of X-ray detectors consisting of tetrahedral units. X-ray technologies are well developed and overall a lot less expensive.

A summary of configurations is presented in Table \ref{tab:summary}, where the event rate is estimated as 
\begin{equation}
\label{eq:ER}
{\rm Event~rate}
\simeq\eta\left(\frac{\sigma}{\sigma_0}\right)
\langle\dot{N}\rangle\frac{d^2}{R_\oplus^2}\,,
\end{equation}
similar to Ref. \cite{Liang:2019lya}, with a more sophisticated estimation based on the Monte Carlo evaluation of cross section. Here $\eta\in(0,1)$ is the efficiency of the detection and is assumed to be of order 1. $\sigma$ and $\sigma_0$ are the effective cross section of the network and a single station, respectively (see Appendix \ref{subapp:Cross section of certain network configurations} for details). $R_\oplus=6371$ km is the radius of Earth.

In what follows, we investigate the optimal configurations for axions and X-rays.

\begin{table*}[!htp] 
	\captionsetup{justification=raggedright}
	\caption{Synchronized networks. The incoming AQN flux is assumed to be isotropic. For axions (resp. X-rays), the nearest distance between detectors is chosen to be 100 km (resp. 50 m). The typical time delay $\Delta t$ and bandwidth $\tau$ are of order 0.1 seconds (resp. 0.1 milliseconds). 
    The efficiency parameter $\eta$ is chosen to be 0.2. 
	} 
	\centering 
	\begin{tabular}{ccccccc}
		\hline \hline
		Radiation & Configuration  & $\Delta t$ [s] & $\tau$ [s] & $\sigma/\sigma_0$  & Event rate [${\rm day}^{-1}$] & Specifications 
		\\  \hline  
		---  & Single detector  & --- &  ---  & 1  & 2.86   & no constraint \eqref{eq:bold hat z cdot bold d_i} applied   \\ 
		axion &  Equilateral triangle & $0.22$  & $0.24$    & 0.12  & 0.33 & 94\% unique solutions    \\ 
		axion & Square & $0.17$ & $0.22$   & 0.14  & 0.39 & ---   \\
		X-ray & Regular tetrahedron & $1.03\times10^{-4}$ & $1.08\times10^{-4}$  & 0.12 & $8.4\times10^{-8}$ & ground-based   \\ 
		X-ray & Regular tetrahedron & $9.48\times10^{-5}$ & $1.15\times10^{-4}$   & 0.35  & $2.5\times10^{-7}$ & 50 m above ground   \\ 
		X-ray & X-ray arrays $(10\rm\,km)^2$ & $1.14\times10^{-4}$ & $1.37\times10^{-4}$  & 0.64   & $5.8\times10^{-3}$  & $10^5$ X-ray detectors  \\ 
		X-ray & X-ray arrays $(100\rm\,km)^2$ & $1.14\times10^{-4}$ & $1.37\times10^{-4}$  & 0.64  & 0.58  & $10^7$ X-ray detectors  \\ 
		\hline\hline
	\end{tabular}
	\label{tab:summary}
\end{table*}

\subsubsection{Axion: network of broadband stations}
As discussed in subsection \ref{subsec:Reconstruction from a synchronized network}, the trajectory \eqref{eq:bold r(t)}  can be identified by synchronized measurements from three stations, based on set of Eqs. \eqref{eqs:bold d_1 cdot bold v etc}. In Appendix \ref{app:Degenerate solutions in Eqs.}, it is proven that Eqs. \eqref{eqs:bold d_1 cdot bold v etc} yields four degenerate solutions. To correctly identify the true trajectory out of the four, one should either introduce additional physical constraints or add stations to the network. 

\begin{figure}[h]
	\centering
	\begin{subfigure}[b]{0.23\textwidth}
		\centering
		\includegraphics[width=\textwidth]{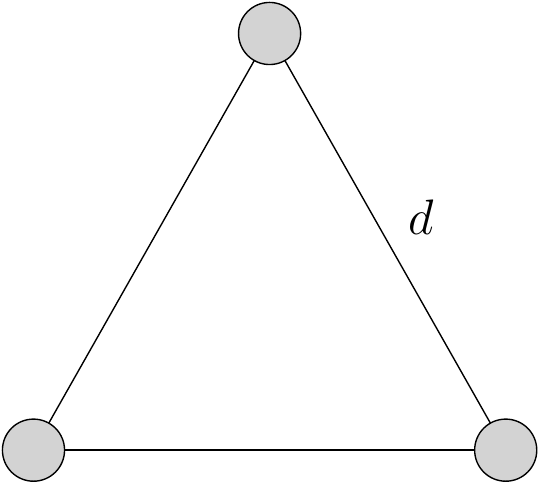}
		\caption{}
		\label{fig:config_triangle}
	\end{subfigure}
	\hfill
	\begin{subfigure}[b]{0.21\textwidth}
		\centering
		\includegraphics[width=\textwidth]{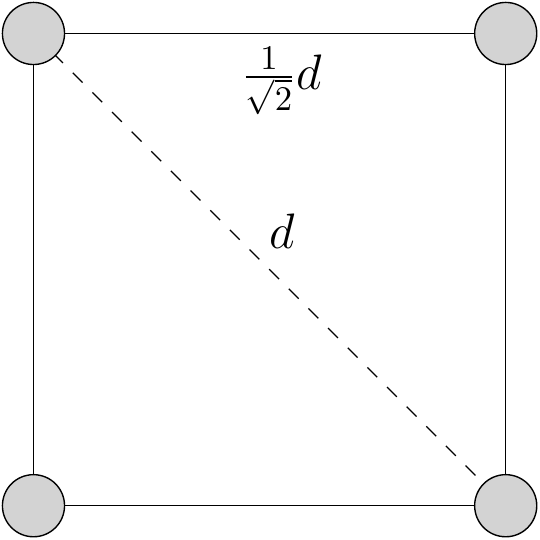}
		\caption{}
		\label{fig:config_square}
	\end{subfigure}
	\caption{Two network configurations for the detection of axions: (a) An equilateral triangle, and (b) a square network. Optimally, stations (grey circle) are separated by a distance $d\simeq100$ km. 94\% (resp. 100\%) of the trajectories are uniquely determined in a triangular (resp. square) network. A triangular network has the highest event rate per station.}
	\label{fig:config_axion}
\end{figure}

When antimatter AQNs annihilate with normal matter inside Earth, there is an additional constraint: axion emission only takes place underground. This implies three additional constraints for axions:
\begin{equation}
\label{eq:bold hat z cdot bold d_i}
\mathbf{\hat{z}}\cdot\mathbf{d}_i<0 \quad
(i=1,2,3)\,
\end{equation}
Simulation indicates that  \eqref{eq:bold hat z cdot bold d_i} is sufficient to ensure 94\% of the trajectories are uniquely determined by Eqs. \eqref{eqs:bold d_1 cdot bold v etc} (see Appendix \ref{subapp:Uniqueness of solutions} for details). The optimal configuration is a three-station network: an equilateral triangle with  $d\simeq100$ km (Fig. \ref{fig:config_triangle}). The event rate is $1.7\eta\rm\,day^{-1}$.

An alternative way to eliminate the degeneracy is to construct a four-station network, with no additional constrains (see Appendix \ref{subapp:Uniqueness of solutions}). Fig. \ref{fig:config_square} illustrates the optimal configuration of such a network: each station is located at the vertices of a square, with $d\simeq 100$ km. The trade off is the limited improvement in event rate ($2.0\eta\rm\,day^{-1}$) compared to the triangular network ( $1.7\eta\rm\,day^{-1}$). 
If uniqueness is not a major concern,  it is more efficient to enhance the event rate by extending a triangular network  than using a square one. 

\subsubsection{X-rays: array of ground-based detectors}
\begin{figure}[h]
	\centering
	\includegraphics[width=0.5\linewidth]{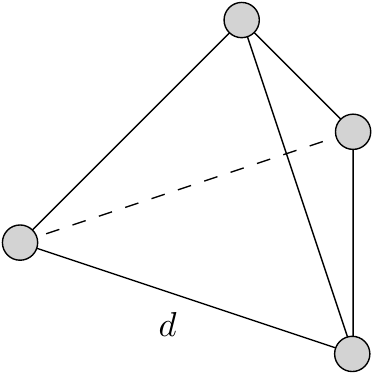}
	\caption{A base unit of the X-ray array of detectors. It consists of four detectors (grey circles) located at the vertices of a regular tetrahedral, separated by $d=50$ m. }
	\label{fig:config_tetrahedron}
\end{figure}

X-rays propagate a shorter distance in air ($d\simeq50$ m) than axions. It is therefore possible to construct a regular tetrahedral network ( Fig. \ref{fig:config_tetrahedron}). A tetrahedral network is optimal as it has both a high event rate (due to its triangular structure) and a unique solution (it consists of four detectors). It can serve as a base unit to form a large array of detectors. For increased performance, the array should be lifted 50 meters above ground so that the detection covers all directions in space. 

According to simulation in Appendix \ref{subapp:Cross section of certain network configurations}, the event rate for a single tetrahedral unit is $4.2\times10^{-7}\eta\rm\,day^{-1}$ (ground based) and $1.2\times10^{-6}\eta\rm\,day^{-1}$ (50 m above ground). For an array covering a surface area by $(100\rm\,km)^2$, the event rate is estimated to be $2.9\eta\rm\,day^{-1}$.

\section{Conclusion}
\label{conclusion}
The main goal of this work is to develop a new search strategy which allows to  study the directionality and velocity distribution of DM particles. To reach this goal, we suggest using a network of synchronized  detectors and we apply this idea to two specific cases: 1.   a broadband detection strategy  to search for axions, within the context of the Axion Quark Nugget (AQN) dark matter  model; 2. a network of X-ray detectors that allows to reconstruct the velocity vector of incident AQN particles. We argue that by looking at the time delays recorded by two or more nearby detectors,  such a network  allows to discriminate between the DM signal and larger noise.

 In case 1, we focused on typical ``small" events, i.e. AQN with a baryon charge  $B  \sim 10^{25}$ , propagating through the Earth and its atmosphere once a day per $(100\,\rm km)^2$. We showed that a network of four synchronized detectors is necessary to uniquely determine the directionality of the incident flux and the velocity distribution of the DM flux. On the other hand, we showed that an optimal configuration for a direct detection of AQN-induced axions, when AQNs penetrate the Earth, is a synchronized network of three stations placed at the vertices of an equilateral triangle, 100 km apart. For an array covering a surface area of $(100\rm\,km)^2$, the event rate is estimated to be $1.7\eta\rm\,day^{-1}$. Such a network uniquely determine the directionality of the incident DM particle in 94\% of cases.

AQNs propagating through the Earth atmosphere are accompanied by signature X-rays, which therefore constitute an indirect detection of AQNs.  X-ray stations are easier to build and operate then axion detectors. In case 2, we showed that the optimal configuration to detect this emission is an array of tetrahedral units. Each unit consist in four synchronised detectors positioned at the vertices of a regular tetrahedron, 50 m apart. Such a network allows to uniquely determine the directionality of the emission. For an array covering a surface area of $(100\rm\,km)^2$, the event rate is estimated to be $2.9\eta\rm\,day^{-1}$. One should emphasize that such a network can eliminate the dominant portion of spurious signals. Therefore, recording synchronized events would be a very strong argument supporting their unconventional nature \footnote{It would be very interesting to study if  the  observed  stratospheric temperature  anomalies being interpreted  in terms of the ``invisible matter stream" \cite{Zioutas:2020ndf}   are somehow related to the excess of energy deposited by the AQNs into atmosphere.}.

Detecting DM particles with those networks can have fundamental consequences since reconstructing the directionality and velocity distribution of DM particles leads to testing the Standard Halo Model.
\section*{Acknowledgments}
AZ is thankful to D. Budker and V. Flambaum for infinitely long, never ending discussions about possible networks, which motivated the present studies. This work was supported in part by the National Science and Engineering Research Council of Canada and
X.L. by the UBC four year doctoral fellowship. We are grateful to Benoite Pfeiffer for a careful reading of the manuscript.

\appendix

\section{Degenerate solutions in Eqs. \eqref{eqs:bold d_1 cdot bold v etc}}
\label{app:Degenerate solutions in Eqs.}
To demonstrate the existence of degenerate solutions, we first solve for Eqs. \eqref{eqs:bold d_1 cdot bold v etc} perturbatively in the following limit of mirror symmetry (i.e., $2\leftrightarrow3$ with similar values):
\begin{subequations}
	\begin{equation}
	\label{eq:t_3}
	t_3=(1+\varepsilon_t)t_2,
	\quad \tau_3=(1+\varepsilon_\tau)\tau_2,
	\end{equation}
	\begin{equation}
	\label{eq:bold R_2}
	\mathbf{R}
	=R(\mathbf{\hat{x}}\cos\gamma
	-\mathbf{\hat{y}}\sin\gamma),\quad
	\mathbf{R}_3
	=R(\mathbf{\hat{x}}\cos\gamma+\mathbf{\hat{y}}\sin\gamma)\,,
	\end{equation}
\end{subequations}
where $\varepsilon_t,\varepsilon_\tau\ll1$, and $2\gamma$ is the angle between $\mathbf{R}_2$ and $\mathbf{R}_3$. One of the solution is
\begin{subequations}
\label{eqs:bold v etc}
\begin{equation}
\label{eq:bold v}
\mathbf{v}
=\frac{R/\tau_1}{S^2+(\cos\gamma- T)^2}
\begin{pmatrix}
S  \\
\frac{\varepsilon_t}{2\sin\gamma}\frac{t_2}{\tau_1}  \\
\cos\gamma- T
\end{pmatrix},
\end{equation}
\begin{equation}
\label{eq:bold d_1}
\mathbf{d}_1
=\frac{R}{S^2+(\cos\gamma- T)^2}
\begin{pmatrix}
\cos\gamma- T  \\
\frac{-1}{2\sin\gamma}
\frac{t_2^2\varepsilon_t+\tau_2^2\varepsilon_\tau}{\tau_1^2}  \\
-S
\end{pmatrix}\,,
\end{equation}
\end{subequations}
where $S$ and $T$ are dimensionless parameters determined by $t_i,\tau_i$ and $\gamma$:
\begin{subequations}
	\label{eqs:S and T}
	\begin{equation}
	\label{eq:S}
	S=\frac{t_2}{\tau_1\cos\gamma}(1+\frac{1}{2}\varepsilon_t)\,,\quad
	\end{equation}
	\begin{equation}
	\label{eq:T}
	T=\sqrt{
		\frac{t_2^2(1+\varepsilon_t)+\tau_2^2(1+\varepsilon_\tau)}{\tau_1^2}
		-1+\cos^2\gamma-S^2
	}\,.
	\end{equation}
\end{subequations}
The three remaining solutions are obtained by applying one of, or a combination of, the following operators to solution \eqref{eqs:bold v etc}:
\begin{subequations}
	\begin{equation}
	\label{eq:cal R}
	{\cal R}:\quad v_z\rightarrow -v_z,\quad d_{1,z}\rightarrow -d_{1,z}\,;
	\end{equation}
	\begin{equation}
	\label{eq:cal T}
	{\cal T}:\quad T\rightarrow -T\,.
	\end{equation}
\end{subequations}
The reflection ${\cal R}$ is an obvious symmetry in Eqs. \eqref{eqs:bold d_1 cdot bold v etc}, while the modified time reversal ${\cal T}$ is less apparent. Note that in Eqs. \eqref{eqs:bold d_1 cdot bold v etc}, the time reversal cannot be obtained by simply flipping the sign of time $t\rightarrow-t$ because one has to also reverse the labeling of each station $(1,2,3)\rightarrow(3,2,1)$ at the same time. Therefore, to reverse the time without changing the labeling, we will need the modified ``time'' $T$ as defined in Eqs. \eqref{eq:T}, and then we take the time reversal of $T$. 

\begin{figure}[h]
	\centering
	\includegraphics[width=0.9\linewidth]{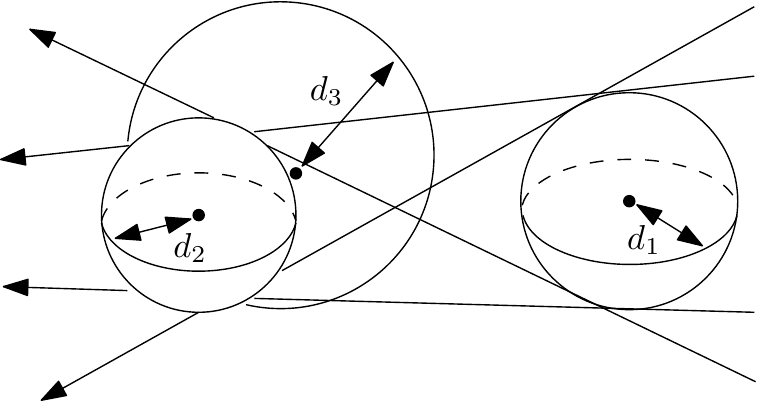}
	\caption{Qualitative picture of the four degenerate solutions ($\mathbb{Z}_2\times\mathbb{Z}_2$) in Eqs. \eqref{eqs:bold d_1 cdot bold v etc}. Allowed trajectories are simultaneously tangent to three spheres of given radii $d_i$ ($i=1,2,3$), where $d_i$ is the distance of the AQN trajectory from the $i$-th detector and is determined by the input parameter $(\tau_i,t_i)$. In practice, the radii change moderately upon the modified time reversal $\cal{T}$ transformation to account for the rescaled $(\mathbf{v},\mathbf{d}_1)$.}
	\label{fig:nonuniqueness}
\end{figure}

We make an intuitive plot of the four degenerate solutions ($\mathbb{Z}_2\times\mathbb{Z}_2$) in Fig. \ref{fig:nonuniqueness}. Qualitatively, solving Eqs. \eqref{eqs:bold d_1 cdot bold v etc} is equivalent to looking for trajectories simultaneously tangent to three sphere of given radii $d_i$ ($i=1,2,3$), where $d_i$ is the distance of the AQN trajectory from the $i$-th detector and is determined by the input parameter $(\tau_i,t_i)$. Apparently, there are always four tangent lines that satisfy such criteria. In practice, ${\cal T}\{d_i\}$ are moderately different from the original $\{d_i\}$ because the transformation accounts for a modified time scale.

\section{Details of numerical simulation}
\label{app:Details of numerical simulation}
The details of all numerical simulations based on Monte Carlo method are presented in this appendix. The distribution of the AQN trajectories follows an isotropic distribution as derived in Ref. \cite{Lawson:2019cvy}, namely
\begin{subequations}
	\label{eqs:cos psi and varphi etc}
	\begin{equation}
	\cos\psi\sim {\rm uniform}(-1,1),\quad
	\varphi\sim{\rm uniform}(0,2\pi)\,,
	\end{equation}
	\begin{equation}
	\label{eq:v sim}
	v
	\sim v^3\frac{\sinh(\mu v/\sigma^2)}{\mu v/\sigma^2}
	\exp\left[-\frac{v^2+\mu^2}{2\sigma^2}\right]\,,
	\end{equation}
	\begin{equation}
	\cos\theta\sim {\rm uniform}(-1,1),\quad
	\phi\sim{\rm uniform}(0,2\pi)\,,
	\end{equation}
	\begin{equation}
	d_1\sim{\rm uniform}(0,d)\,,
	\end{equation}
\end{subequations}
where $\mu=220\kmps$ is mean galactic velocity, $\sigma=110\kmps$ is the galactic dispersion, the effective distance of detection $d$ is 100 km (resp. 50 m) in case of axions (resp. X-rays). Following the simulation conditions \eqref{eqs:cos psi and varphi etc}, we define two angles
\begin{equation}
\begin{aligned}
\mathbf{\hat{n}}
&\equiv(\sin\theta\cos\phi,\sin\theta\sin\phi,\cos\theta)\,,\\
\mathbf{\hat{v}}
&\equiv(\sin\psi\cos\varphi,\sin\psi\sin\varphi,\cos\psi)\,,
\end{aligned}
\end{equation}
so that a trajectory \eqref{eq:bold r(t)} is generated with given $\mathbf{v}$ and $\mathbf{d}_1$:
\begin{equation}
\label{eq:hat_v and hat_d}
\mathbf{v}
=v\mathbf{\hat{v}}\,,\quad
\mathbf{d}_1
=d_1\frac{\mathbf{\hat{n}}\times\mathbf{\hat{v}}}
{|\mathbf{\hat{n}}\times\mathbf{\hat{v}}|}\,.
\end{equation}
Sec. \ref{subsec:Optimal configurations of a synchronized network} suggests the optimal configurations of a synchronized network are equilateral triangle, square, and regular tetrahedron with separation distance of order $d$. For illustrative purpose, we set the distance between two neighboring stations to be $\Delta R=d$ (and $d/\sqrt{2}$ in the square case) in the following simulations.  

One more implicit constraint is that all stations should have the same range of effective distance of detection, namely
\begin{equation}
\label{eq:d_i}
d_i\leq d\,,
\end{equation}
for each $i$-th station. From the definition \eqref{eq:abs bold r(t_i)}, the time input parameters are related to the simulated $\mathbf{v}$ and $\mathbf{d_1}$ in what follows:
\begin{equation}
\label{eqs:t_i and tau_i}
t_i=\frac{\mathbf{v}\cdot\mathbf{R}_i}{v^2}\,,\quad
\tau_i
=\frac{1}{v}\sqrt{
	R_i^2+d_1^2-(\mathbf{\hat{v}}\cdot \mathbf{R}_i)^2
	-2\mathbf{d}_1\cdot\mathbf{R}_i}\,.
\end{equation}

For simulations in this appendix, the trajectories are generated following the above recipe described.

\subsection{Sensitivity to time measurement}
\label{subapp:Sensitivity to time measurement}
\begin{figure}[h]
	\centering
	\includegraphics[width=\linewidth]{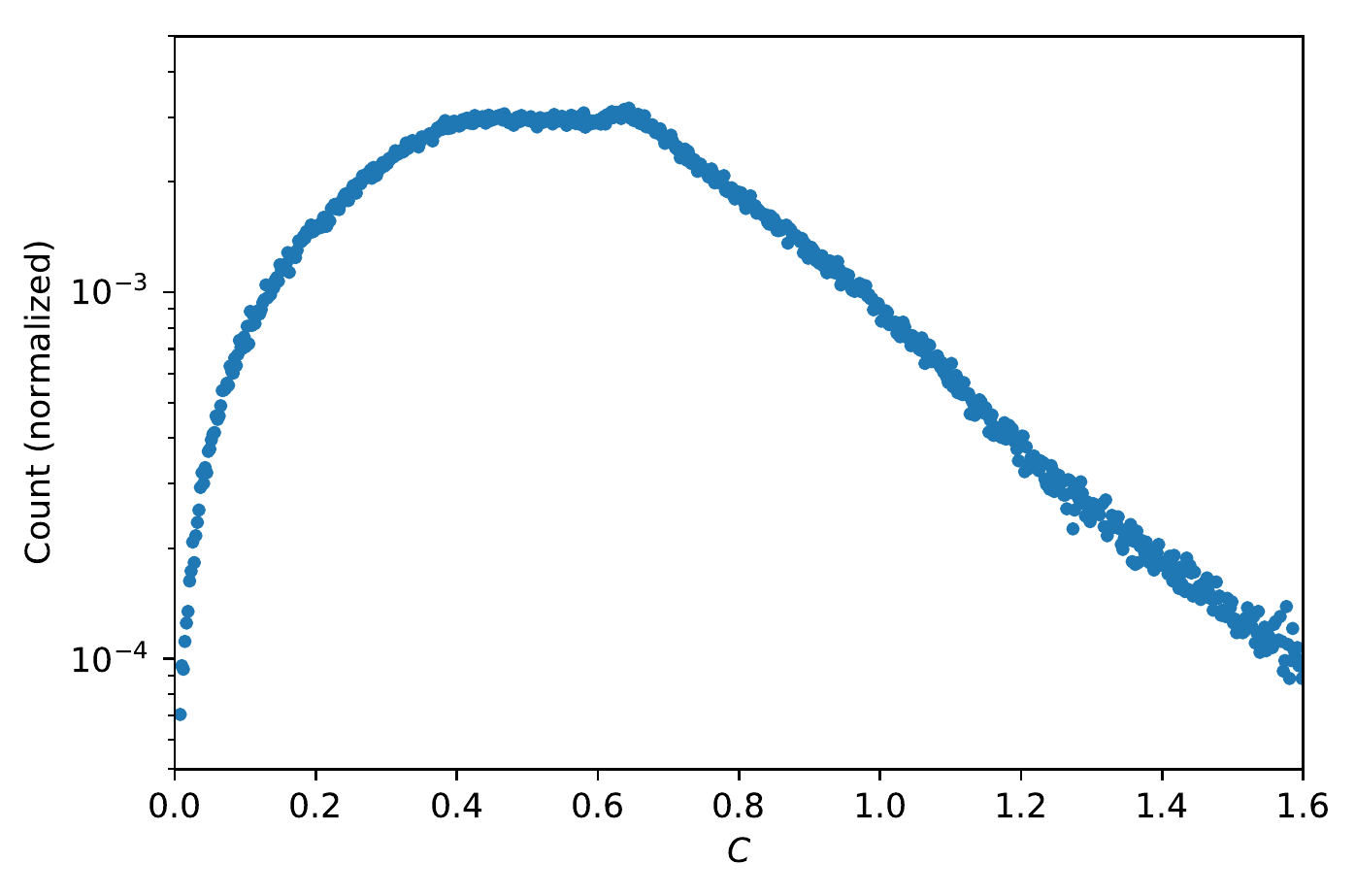}
	\caption{Histogram of $C$ fitting into inequality \eqref{eq:abs delta bold v v} in Monte Carlo simulation ($10^6$ samples). The parameter $\varepsilon_{\rm cut}$ is chosen to be 0.5, and the count is normalized. The value of $C$ is found to be 1.24 (1.92) among 95\% (99\%) of samples.}
	\label{fig:histogram}
\end{figure}

In addition to a simulated trajectory \eqref{eq:bold r(t)} parameterized by $(\mathbf{v},\mathbf{d}_1)$, a random perturbation  $(\delta\mathbf{v},\delta\mathbf{d}_1)$ should be generated:
\begin{subequations}
	\begin{equation}
	\cos\psi'\sim {\rm uniform}(-1,1),\quad
	\varphi'\sim{\rm uniform}(0,2\pi)\,,
	\end{equation}
	\begin{equation}
	\cos\theta'\sim {\rm uniform}(-1,1),\quad
	\phi'\sim{\rm uniform}(0,2\pi)\,,
	\end{equation}
	\begin{equation}
	\frac{\delta v}{v}\sim {\rm uniform}(0,\varepsilon_{\rm cut})\,,
	\end{equation}
\end{subequations}
where $\varepsilon_{\rm cut}$ is a cutoff of perturbation and we choose its upper limit to be 0.5 in the numerical simulation. Now we define
\begin{subequations}
	\begin{equation}
	\begin{aligned}
	\mathbf{\hat{n}}'
	&\equiv(\sin\theta'\cos\phi',\sin\theta'\sin\phi',\cos\theta')\,,\\
	\mathbf{\hat{v}}'
	&\equiv(\sin\psi'\cos\varphi',\sin\psi'\sin\varphi',\cos\psi')\,,
	\end{aligned}
	\end{equation} 
	\begin{equation}
	\delta\mathbf{v}
	=\delta v\,\mathbf{\hat{v}}',\quad
	\delta\mathbf{d}_1
	=\delta d_1\frac{\mathbf{\hat{n}}'\times\mathbf{\hat{v}}'}
	{|\mathbf{\hat{n}}'\times\mathbf{\hat{v}}'|}
	\equiv\delta d_1\mathbf{\hat{d}}'_1\,,
	\end{equation}
\end{subequations}
where $\delta d_1$ is completely determined by the constraint of orthogonality $\mathbf{v}\cdot\mathbf{d}_1=0$, namely
\begin{equation}
\frac{\delta d_1}{d_1}
\equiv-\frac{\delta v}{v}
\frac{\mathbf{\hat{d}}_1\cdot\mathbf{\hat{v}}'}
{\mathbf{\hat{v}}\cdot\mathbf{\hat{d}}_1'}\,.
\end{equation}
Then we obtain 
\begin{equation}
\begin{aligned}
\delta t_i
&\equiv t_i(\mathbf{v}+\delta\mathbf{v})-t_i(\mathbf{v})\,,\\
\delta\tau_i
&\equiv\tau_i(\mathbf{v}+\delta\mathbf{v},\mathbf{d}_1+\delta\mathbf{d}_1)
-\tau_i(\mathbf{v},\mathbf{d}_1)
\end{aligned}
\end{equation}
from Eqs. \eqref{eqs:t_i and tau_i}. Additionally, we only keep data points of which the perturbations are not overly large:
\begin{equation}
\sqrt{\sum_{i=1}^{3}\left(\frac{\delta t_i}{t_i}\right)^2
	+\left(\frac{\delta\tau_i}{\tau_i}\right)^2}
\leq1\,.
\end{equation}
because the uncertainty of time measurement is assumed small. From Monte Carlo simulation, a large sample array of $(\delta t_i,\delta\tau_i,\delta \mathbf{v},\delta \mathbf{d}_1)$ is generated. Fitting the data set into the inequality \eqref{eq:abs delta bold v v}, $C$ is found to be 1.24 (resp. 1.92) among 95\% (resp. 99\%) of $10^6$ samples at $\varepsilon_{\rm cut}=0.5$, see Fig. \ref{fig:histogram}. The value of $C$ is insensitive ($\leq0.1$ absolute uncertainty) to cutoff $\varepsilon_{\rm cut}$ and choice of separation distance $\delta R$.

\begin{figure}
	\centering
	\includegraphics[width=\linewidth]{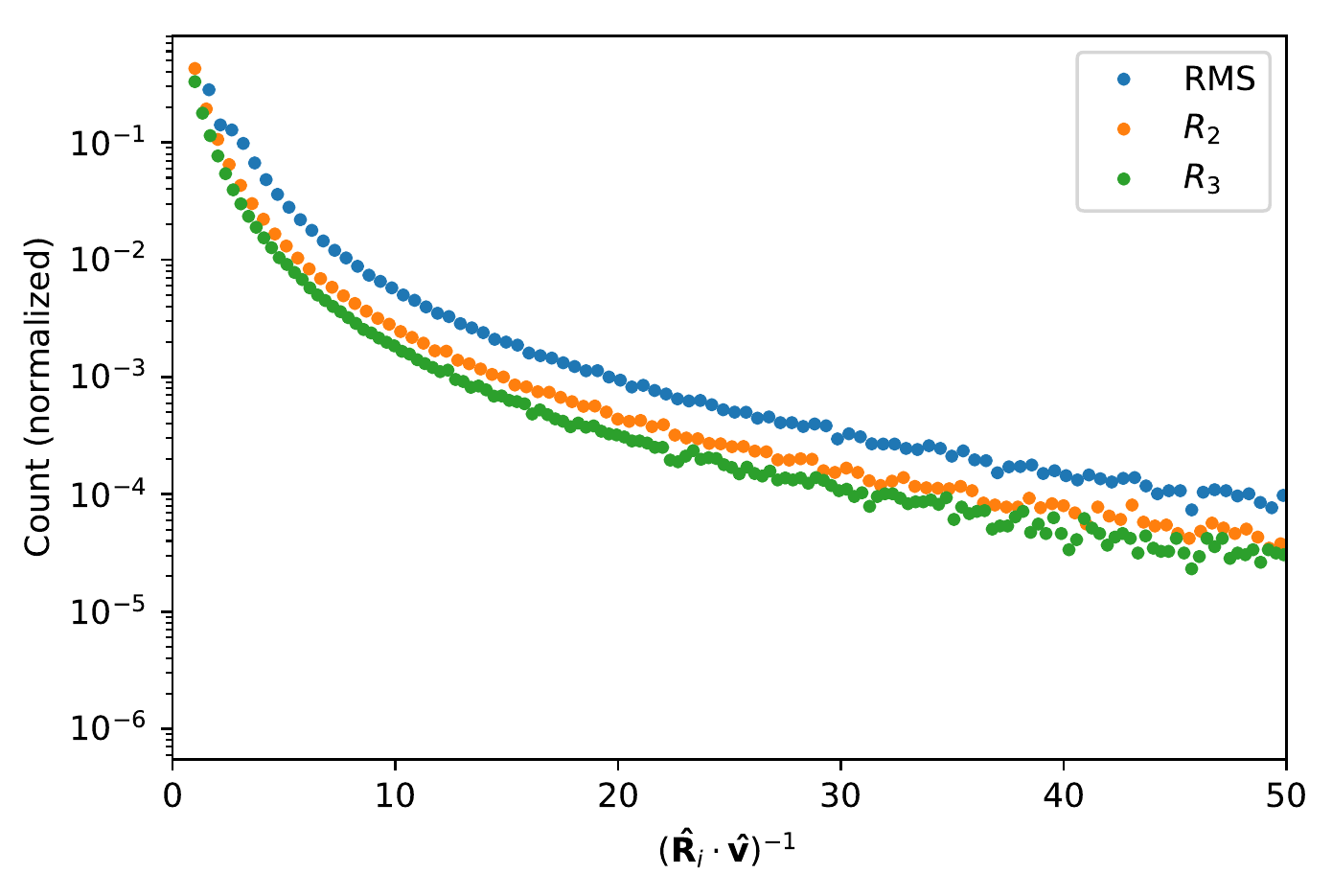}
	\caption{Histogram of $\langle(\mathbf{\hat{R}}\cdot\mathbf{\hat{v}})^{-1}\rangle$ as defined in Eq. \eqref{eq:langle abs delta v v rangle}. Monte Carlo simulation ($10^6$ samples). The count is normalized. The value of $\langle(\mathbf{\hat{R}}\cdot\mathbf{\hat{v}})^{-1}\rangle$ is found to be 12.3 (resp. 32.0) among 95\% (resp. 99\%) of samples.}
	\label{fig:histogramgamma}
\end{figure}

One more important quantity is the root mean square $\langle(\mathbf{\hat{R}}\cdot\mathbf{\hat{v}})^{-1}\rangle$ as defined in Eq. \eqref{eq:langle abs delta v v rangle}. As shown in Fig. \ref{fig:histogramgamma}, the value of $\langle(\mathbf{\hat{R}}\cdot\mathbf{\hat{v}})^{-1}\rangle$ is found to be 12.3 (resp. 32.0) among 95\% (resp. 99\%) of $10^6$ samples, which is independent of $\Delta R$ by its definition.

\subsection{Cross section of certain network configurations}
\label{subapp:Cross section of certain network configurations}
We generate $10^7$ samples using Eqs. \eqref{eqs:cos psi and varphi etc} to Eqs. \eqref{eq:hat_v and hat_d}, with the modification that 
\begin{equation}
d_1\sim{\rm uniform}(0,Nd),
\end{equation}
and then check what percentage of the samples satisfy the condition \eqref{eq:d_i} for each station in different configurations, in the case of axions we also seek to satisfy the constraint \eqref{eq:bold hat z cdot bold d_i}. We also take the ratio $\sigma/\sigma_0$ of each configuration to the single station.
This was repeated for  $N = 1,\, 2, \, 100$ and for $d=1,\,100$ but it was found to have no effect on the percentage of trajectories satisfying the conditions. The configurations used are as described in subsection \ref{subsec:Optimal configurations of a synchronized network} where $\Delta R = d$.
Additionally for X-ray we also used a flat disk configuration, where for each trajectory we see what percentage of samples satisfy $r_z = 0$ and $r_x^2 + r_y^2 < R^2$. Results are summarized in Table \ref{tab:summary}. The cross section ratio $\sigma/\sigma_0$ determines the event rate of a local flash as defined in Eq. \eqref{eq:ER}. We conclude $\sigma/\sigma_0\sim \mathcal{O}(0.1)$ for all configurations listed as expected.

\subsection{Uniqueness of solutions}
\label{subapp:Uniqueness of solutions}
We first perform a simulation check for the axion network. In this case, additional constraints \eqref{eq:bold hat z cdot bold d_i} applies to every station. Both triangular and square configurations are examined. The time delay $t_i$ and bandwidth $\tau_i$ are retrieved from Eqs. \eqref{eqs:t_i and tau_i}. Feeding the simulated time parameters into Eqs. \eqref{eqs:bold d_1 cdot bold v etc}, the equation set is solved under constraints \eqref{eq:bold hat z cdot bold d_i}. For solutions that have no more than 10\% difference in velocity from each other, they are treated as identical. With $10^6$ samples simulated, we find 94\% of trajectories are the unique solutions in Eqs. \eqref{eqs:bold d_1 cdot bold v etc} in the triangular configuration, and the rate of uniqueness is 100\% in the square configuration. 

For the x-ray network,  the tetrahedral configuration is examined under the same process without the constraints \eqref{eq:bold hat z cdot bold d_i} for generality. With $10^6$ samples simulated, we find the rate of uniqueness is 100\%.

\subsection{Time delay and bandwidth}
\label{subapp:Time delay and bandwidth}
\begin{figure}[h]
    \includegraphics[width=1\linewidth]{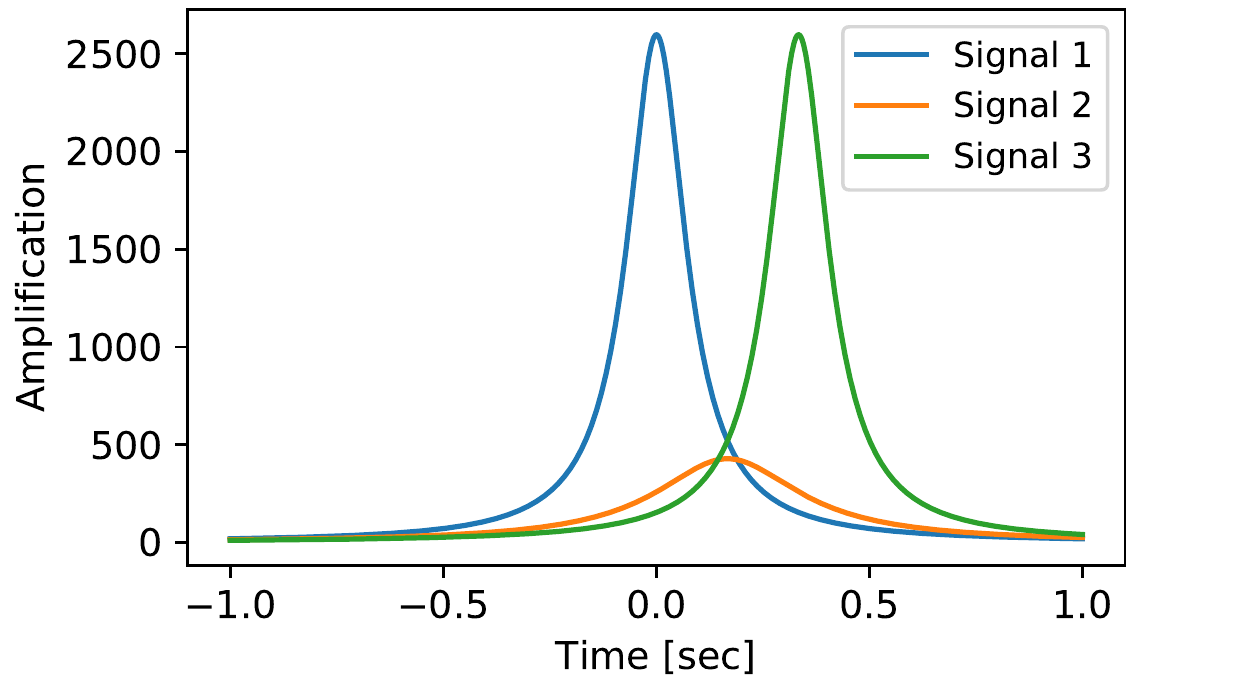}
    \caption{Signals generated for a triangular configuration. Each signal corresponds to each station detecting the axion flux due to a nearby AQN. From these signals we can obtain the parameters $t_i$ and $\tau_i$ in order to solve the system of Eqs. \eqref{eqs:bold d_1 cdot bold v etc} }
    \label{fig:triangle signals}
\end{figure}

In order to get the typical values of time delay and bandwidth we do a Monte Carlo simulation with $10^7$ samples. During the simulation for each valid trajectory we record the values of $t_i$ and $\tau_i$ of each station according to Eqs. \eqref{eqs:t_i and tau_i}, and then we take the average value along with the standard deviation. The statistics for $t_1$ and $\tau_1$ are not collected because $t_1$ is always $0$ by setting \eqref{eq:bold R_1}. A sample simulated signal received in a triangular network is presented in Fig. \ref{fig:triangle signals}. The average time delay $|\Delta t|$ and average bandwidth share similar values:  0.2 s (resp. 0.1 ms) for radiation of axions (resp. X-rays). This is as expected from definitions \eqref{eq:t_i-t_j} and \eqref{eq:tau}. The results are summarized in Table \ref{tab:summary}. 

\FloatBarrier

\appendix
 
\bibliography{Network}

\end{document}